# Ontology-based Approach for Identifying the Credibility Domain in Social Big Data

*Pornpit Wongthontham*[1] *and Bilal Abu-Salih*[2]

*Curtin University, GPO Box U1987 Perth WA 6845 AUSTRALIA,*

[1]*p.wongthongtham@curtin.edu.au,* [2]*bilal.abusalih@curtin.edu.au*

**Abstract**

The challenge of managing and extracting useful knowledge from social media data sources has attracted much attention from academics and industry. To address this challenge, semantic analysis of textual data is focused on in this paper. We propose an ontology-based approach to extract semantics of textual data and define the domain of data. In other words, we semantically analyse the social data at two levels i.e. the entity level and the domain level. We have chosen Twitter as a social channel for the purpose of concept proof. Ontologies are used to capture domain knowledge and to enrich the semantics of tweets, by providing specific conceptual representation of entities that appear in the tweets. Case studies are used to demonstrate this approach. We experiment and evaluate our proposed approach with a public dataset collected from Twitter and from the politics domain. The ontology-based approach leverages entity extraction and concept mappings in terms of quantity and accuracy of concept identification.

## 1. Introduction

Nowadays, we are surrounded by a large volume of data and information from a multitude of sources. Data has been generated at approximately 2.5 exabytes a day (IBM, 2015). It is a huge challenge to manage and extract useful knowledge from a large quantity of data given the different forms of data, streaming data and uncertainty of data. 'Big data' is recently termed and is a popular phenomenon not only about storage or access to data but also data analytics aiming to make sense of data and to obtain value from data. Big data is defined through the 5V model i.e. Volume, Velocity, Variety, Value, and Veracity (Hitzler P & K., 2013). Within its description, Big Data provides a wealth of information that businesses, political governments, organisations, etc. can mine and analyse to exploit value in a variety of areas. However, there are still challenges in this area of Big Data Analytics research to capture, store, process, visualise, query, and manipulate datasets to develop meaningful information that is specific to an application's domains.

Being able to discover and understand data is a goal of enterprises today. The rapid increase in the amount of unstructured data has highlighted the importance of such data as a means of acquiring deeper and more accurate insights into businesses and customers. These insights here achieve a competitive advantage in the current competitive environment. According to the International Data Corporation (IDC), unstructured data accounts for 80% of the total data in organizations (Gantz & Reinsel., 2010). The amount of unstructured data is expected to increase by 60% per year in the next few years (Gantz & Reinsel., 2010). The emergence of social media has contributed to the increase of unstructured data. Social media has given everyone a place to express and share their opinions and thoughts on all kinds of topics. Social media offers a data source for relevant big data which includes shared content, picture, videos, etc. From Twitter's record[1], there are 500 million tweets sent per day and 288 million monthly active users. The vast amount of social data has spread to many different areas in everyday life e.g. e-commerce (Kaplan & Haenlein, 2010), education (Tess, 2013), health (Salathé, Vu, Khandelwal, & Hunter, 2013), etc.

Social big data analysis involves joining two domains: social media and big data analysis. Bello-Orgaz et al. (Bello-Orgaz, Jung, & Camacho, 2016) define the concept of social big data analysis as follows:

*"Those processes and methods that are designed to provide sensitive and relevant knowledge from social media data sources to any user or company from social media data sources when data source*

[1] https://about.twitter.com/company

*can be characterised by their different formats and contents, their very large size, and the online or streamed generation of information." (p. 46)*

One of the biggest challenges is to distinguish the credible information from that which is not. Due to the open environment and few restrictions associated with social media, rumours can spread quickly and false information can be broadcast rapidly. This could impact badly on businesses, political managements, public health, etc. if the false information is being published amongst the trustworthy information. However, if it is accurate information, this could be greatly beneficial to individuals and organisations as a means of acquiring value from social media data. Hence, it is essential to distinguish the trustworthiness to determine the reputation of the sources and to define the legitimate users. A degree of trustworthiness for the data, the sources, and the users is important.

In order to gain insight from social data analytics, in this paper we focus on the semantic analysis of textual data. We propose an ontology-based approach to extract the semantics of textual data and define the credibility domain of data. The credibility domain is the area of knowledge to which extracted information pertains. Information is credible within the boundaries of this domain of knowledge. For example, tweets about politics can be mistakenly classified outside the boundaries of politics domain because the word 'Labor' appears many times in the tweets. However, the word 'Labor' in these tweets refers to a political party hence it should be classified in the boundaries of politics domain.

We aim to semantically analyse the social data at two levels: the domain level and entity level. Due to the typically short length, informality, and irregular structure of messages, we have chosen Twitter as challenge over a social channel in our approach. Twitter[2] is a microblogging platform

[2] https://twitter.com/

where users read and write short messages on various topics every day. We choose politics as the domain because this topic can generate a huge amount of data.

In this work, the domain knowledge is captured in ontologies and we use ontologies to enrich the semantics of tweets provided with specific semantic conceptual representation of entities appearing in tweets. For example, in the politics domain, 'Labor', that is extracted from tweets, would be represented under the concept 'political party' but would be a different concept in another domain such as the Work domain.

In this paper, we present work in progress with optimistic results. We experiment and evaluate our proposed approach with public datasets collected from Twitter and within the politics domain. We evaluate open API tools for concept identification and compare our results with Alchemy[3] from IBM's Watson which, it is claimed by (Rizzo & Troncy, 2011) and (Saif, He, & Alani, 2012), performs best in terms of quantity and accuracy of the identified concepts. The findings conclude that by combining our approach with Alchemy results, the accuracy of concept identification is improved significantly.

The main research question for this paper is as follow:

- How can we identify domain-based credibility in unstructured big data extraction?

In order to answer the above research question, we explore

- Ontology, Linked data, and a Knowledge Base to be used to identify, annotate, and enrich entities in unstructured data,
- the system components when applies in a particular domain i.e. Politics domain, and
- the ontology based approach incorporated with well-known IBM's Watson Alchemy to enhance information extraction.

[3] https://www.ibm.com/watson/alchemy-api.html

The rest of the paper is organised as follows. Section 2 provides background information along with a review of literature relevant to ontology in social media, big social data, and business intelligence in the era of big social data. Section 3 presents the system architecture to semantically analyse tweets. Section 4 describes system components to develop a system in Politics domain. Section 5 provides case studies in politics Twitter data to analyse Politics Twitter data in the case studies. Section 6 provides performance evaluation. Section 7 discusses future research directions in the area. We conclude the paper and discuss future work in Section 8.

## 2. Literature Review

### 2.1 Semantics Analysis

In the latter part of the 20th century, researchers in the field of Artificial Intelligence have become active in computational modelling of ontologies that would deliver automated reasoning capabilities. Tom Gruber generated expansive interest across the computer science community by defining ontology as “an explicit specification of a conceptualisation" (Gruber, 1993). The conceptualisation is the formulating of knowledge about entities. The specification is the representation of the conceptualisations in a concrete form (Stevens, 2001). The specification will lead to commitment in semantic structure. In short, an ontology is the working model of entities. Notably there is development of new software tools to facilitate ontology engineering. Ontology engineering is an effort to formulate an exhaustive and rigorous conceptual schema within a given domain. Basically, ontology captures the domain knowledge through the defined concrete concepts (representing a set of entities), constraints, and the relationship between concepts, in order to provide a formal representation in machine understandable semantics. The purpose of ontology is to represent, share, and reuse existing domain knowledge.

The use of ontology in the social media has been applied widely to infer semantic data in a broad range of applications. Carrasco et al. (Carrasco, Oliveira, Lisboa Filho, & Moreira, 2014) presented an ontology-based, multi-agent solution for the wild animal traffic problem in Brazil. Iwanaga et al. (Iwanaga et al., 2011) and Ghahremanlou et al. (Ghahremanlou, Sherchan, & Thom, 2014) both applied ontology to build applications in crisis situations. The former designed ontology for earthquake evacuation to help people find evacuation centres in earthquakes crises based on data posted in Twitter. The latter showed a geo-tagger that aims to process unstructured content and infer locations with the help of existing ontologies. Bontcheva and Rout (Bontcheva & Rout, 2012) conducted a survey that addressed research issues related to processing social media streams using semantic analysis. Some of the key questions which were the focus of this paper included: (i) How could Ontologies be utilized with Web of Data for semantically annotating social media contents? (ii) How could the annotation process discover hidden semantics in social media? (iii) How could trustworthiness of data be extracted from massive and noisy data? (iv) What are the techniques to model user identity in the digital world? (v) How could information retrieval techniques incorporate semantic analysis to retrieve highly relevant information? Maalej et al. (Maalej, Mtibaa, & Gargouri, 2014) presented an approach that helps mobile users in their search in the social networks by building an ontology-based context-aware module for mobile social networks. Their approach includes: 1. knowledge extraction from SN (implicit, explicit, (none) contextual data using API; 2. data cleansing; 3. knowledge modeling (knowledge of user's details and contextual information); 4. comparing user profiles and the contextual information; and 5. presenting retrieved data in mobile format. Narayan et al. (Narayan, Prodanovic, Elahi, & Bogart, 2010) proposed an approach intended to explore events from a twitter platform and enrich an ontology designed for that purpose.

Statistical techniques have been used as another means of topic modelling and discovery in twitter mining. The two dominant statistical techniques that have been used are LDA (Latent Dirichlet Allocation) (Blei, Ng, & Jordan, 2003), and Latent Semantic Analysis (LSA). In LSA, an early topic modelling method has been extended to pLSA (Hofmann, 1999), which generates the semantic relationships based on a word-document co-occurrence matrix. LDA is based on an unsupervised learning model in order to identify topics from the distribution of words. These approaches have been widely used in sentiment analysis (Saif, He, & Alani, 2011) and several modelling applications (Asharaf & Alessandro, 2015; Li, Wang, Zhang, Sun, & Ma, 2016; Nichols, 2014; Quercia, Askham, & Crowcroft, 2012; Weng, Lim, Jiang, & He, 2010). However the high-level topics classifications that use these bag-of-words statistical techniques are inadequate and inferior (Michelson & Macskassy, 2010). The brevity and ambiguity of such short texts makes more difficult to process topic modelling using these statistical models (Li et al., 2016).

### 2.2 Social Business Intelligence

Berlanga et al. (Berlanga, Aramburu, Llidó, & García-Moya, 2014) suggested a new Semantic Data Infrastructure for the new generation of BI in order to handle the massive amount of unstructured data, and to integrate social media facts and their dimensions into the business intelligence environment. Their first proposed prototype was presented in Garcia-Moya et al. (García-Moya, Kudama, Aramburu, & Berlanga, 2013), where the authors present a methodology and a prototype for processing sentiment data (opinions data, customer review, etc.) and extracting features from such data in order to enhance the data warehouse with new social facts that help to create the new generation of BI. Other work done by Louati et al. (Louati, El Haddad, & Pinson, 2014) addressed the importance of the Voice of Customers (VoC) as a new dimension of BI analytics. Data warehouses and OLAP have been interestingly targeted by researchers to provide more advanced solutions by integrating ontology and semantic web technology. In this context, Albanese (Albanese, 2013) presented a new computational intensive OLAP model to answer semantics queries. Zhang et al. (Zhang, Hu, Chen, & Moore, 2013) illustrated MUSING which allows the extraction of data from heterogeneous sources and uses ontology to annotate extracted information to enhance data quality and provide meaningful information to be employed for the BI goals. The above researches propose solutions in general domains, not specific to any particular domain.

### 2.3 Big Data Analytics

Big Data (BD) technology for data storage and analysis provides advanced technical capabilities to the process of analysing massive and extensive data in order to achieve deep insights in an efficient and scalable manner. Bello-Orgaz et al. (Bello-Orgaz et al., 2016) explored social big data methodologies, social data analytic methods and algorithms, and social based applications. Chen et al. (Y. Chen et al., 2016) conducted an extensive review on big data research focused specially on technological issues. Manyika et al. (Manyika et al., 2011) listed some of the Big Data technologies such as Big Table, Cassandra (Open Source DBMS), Cloud Computing, Hadoop (Open Source framework for processing large sets of data), etc. Although MapReduce and its open source platform (Hadoop) show a robust paradigm to analysis data in the BD context, recent research has focused on dealing with the weaknesses of such a framework and providing alternative solutions such as that of Jiang et al. (Jiang, Chen, Ooi, Tan, & Wu, 2014). Cuesta et al. (Cuesta, Martínez-Prieto, & Fernández, 2013) proposed an architecture to address the Big Semantic Data requirements that take into consideration the in-motion nature (real-time) of the data. Chen et al. (M. Chen, Mao, Zhang, & Leung, 2014) discussed the various open issues and challenges of BD, and listed its key technologies.

The incorporation of BD technology to extend BI tools is considered to be a hot topic, especially within social media because of its significance to BI analytics. This has interestingly attracted researchers in industry and academia to leverage BD techniques to benefit BI tools. Shroff et al. (Shroff, Dey, & Agarwal, 2013) showed three use-cases where social-contents affect BI dramatically: (i) Supply-Chain Disruptions, (ii) VoC, and (iii) Competitive Intelligence. The decision to incorporate BD technology (i.e. Hadoop/MapReduce) for this research is due to the fact that social media content is huge and needs an efficient and scalable technology to manage it, so that the data volume dimension is properly addressed.

Moreover, recent literature has considered Social Networks as a form of Big Data in terms of volume (billions of social links), velocity (massive amount of generated content), and variety (videos, posts , mobile tweets, etc.) (Paik, Tanaka, Ohashi, & Wuhui, 2013). Lim et al. (Lim, Chen, & Chen, 2013),

Cuzzocrea et al. (Cuzzocrea, Bellatreche, & Song, 2013), Shroff et al. (Shroff et al., 2013) and Chen et al. (H. Chen, Chiang, & Storey, 2012) listed the main directions for BI over BD. These include and are not limited to: incorporating BD technology to benefit BI tools, developing methods to handle data in motion (real-time) for BI analysis, designing OLAP tools capable of processing BD, etc.

In addition, we argue that starting from the characteristics of BD and sorting out issues related to these dimensions will be the most efficient way to address BD as well as benefit BI with the expected outcomes of BD Analysis. Saha and Srivastava (Saha & Srivastava, 2014) presented a summary to address the data veracity issue. Poor data quality has a major negative impact on the data analysis process, and the output will lack credibility and trustworthiness. The paper addresses the data quality issues and provides tools and solutions for data in various forms (relational, structured and semi-structured); however, the unstructured data types were not addressed. Moreover, hybrid approaches could be used that utilize ontology for data quality and trust inference purposes. Optique (Calvanese et al., 2013), which is the next generation of Ontology Based Data Access (OBDA), addresses BD characteristics and data access problem in particular. Moreover, Hoppe et al. (Hoppe, Nicolle, & Roxin, 2013) proposed an ontology-based approach for user profiling in the BD context. Reddy (Reddy, 2013) presented a future research project in the distributed semantic data management. The project is divided into two main parts: 1. Design of an actor-based approach paradigm for storing and execution RDF Data over distributed environment utilizing the MapReduce Framework. 2. Proposal of a pay-as-you-go approach for providing Semantic OWL data as a service in the cloud infrastructure; this includes data cleansing and ontologies construction and alignment using the Hadoop/MapReduce platform. In summary, the review of the literature indicates that data analytics for unstructured data is still a challenging area in the context of Big Data.

Table 1 summarises the literature review showing (i) the level of semantics analysis, (ii) whether it makes use of ontology, and (iii) whether it applies in Online Social Networks (OSNs) against the proposed approach that intends to bridge the gap in the literature.

Table 1: A review of selected papers

| **Approach/ Model / Authors** | **Brief Description** | **Semantics Analysis** | | **Use of Ontology** | **Applied in OSNs** |
|---|---|---|---|---|---|
| | | Entity Level | Domain Level | | |
| WATES (Carrasco et al., 2014) | Ontology-based solution for wild animal traffic problem in Brazil. | Yes | No | Yes | Yes |
| Evacuation Ontology (Iwanaga et al., 2011) | An Ontology for earthquake-evacuation for a real-time solution that provides people searching evacuation centers. | Yes | No | Yes | Yes |
| OZCT (Ghahremanlou et al., 2014) | Identifying geographic events by referencing geolocation in tweets. | Yes | No | Yes | Yes |
| (Bontcheva and Rout 2012) | Addressing research issues related to processing social media streams using semantic analysis. | Yes | No | Yes | Yes |
| (Saif et al., 2011) | Sentiment analysis for twitter. | Yes | No | No | Yes |
| TweetLDA (Quercia et al., 2012) | A new supervised topic model for assigning "topics" to a collection of documents. | No | Yes | No | Yes |
| Twitterrank (Weng et al., | Aim to find topic influential twitterers. | No | Yes | No | Yes |
| 2010) | | | | | |
| (Berlanga et al., 2014), (García-Moya et al., 2013), (Louati et al., 2014) | New infrastructure for Social BI. | Yes | No | No | Yes |
| (Albanese, 2013) | To access, retrieve and reuse semantic OLAP databases effectively and efficiently. | Yes | No | Yes | No |
| Epic (Jiang et al. 2014) | Capable, efficient and reliable system to handle data variety well. | No | No | No | Yes |
| SOLID (Cuesta, Martínez-Prieto, and Fernández 2013) | Answer Big Data requirements considering the data that is in-motion nature (real-time). | Yes | No | Yes | No |
| Optique (Calvanese et al., 2013) | Address Big Data characteristics and data access problem in particular. | Yes | No | Yes | No |
| (Hoppe et al., 2013) | Explore an Ontology-based approach for user profiling. | No | Yes | Yes | No |
| (Reddy, 2013) | Distributed semantic data management over cloud based infrastructure. | Yes | No | Yes | No |
| *The proposed* | *Ontology-based approach to extract* | *Yes* | *Yes* | *Yes* | *Yes* |
| *approach* | *semantics of textual data and define the domain of data.* | | | | |

### 3. System Architecture

In this paper, we aim to semantically analyse tweets in order to enrich data with a specific semantic conceptual representation of entities. Essentially, the proposed system has five main processes shown in Figure 1 as follows:

1. Pre-processing data
2. Domain knowledge inference
3. Annotation and enrichment
4. Interlinking entities
5. Semantic Repository

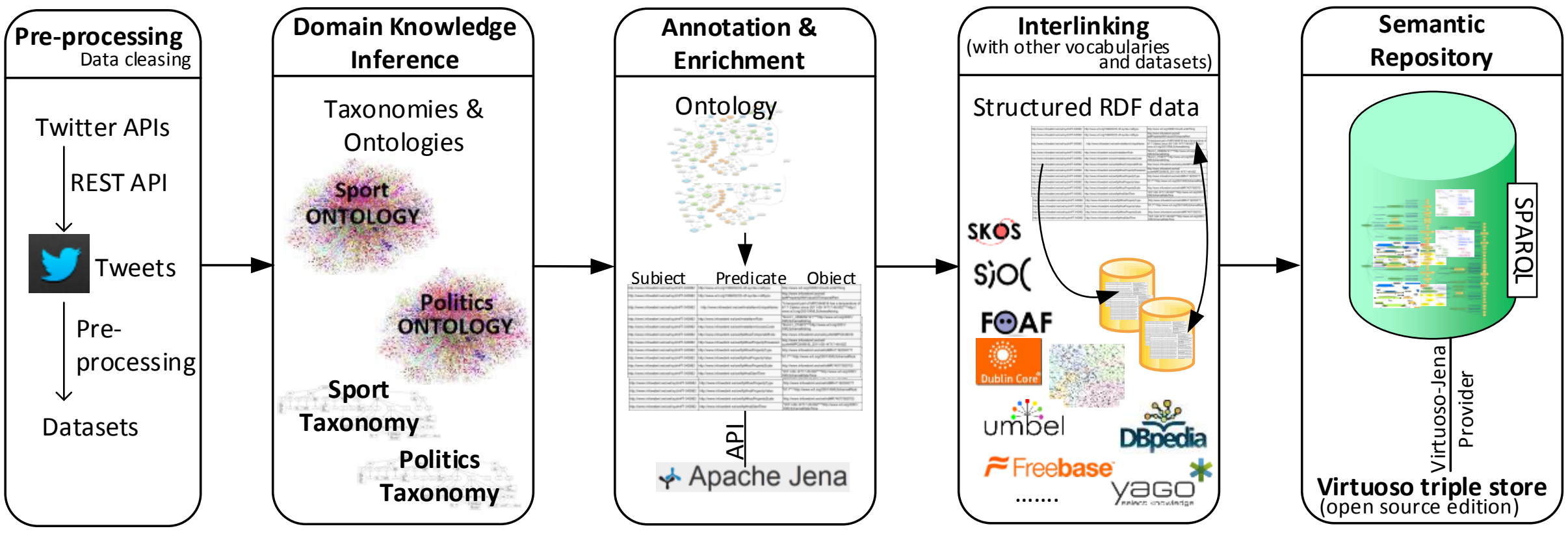

Figure 1: System architecture

### 3.1 Pre-processing data (data cleansing)

We use one of Twitter APIs named REST APIs to collect public archived tweets. The collected tweets are processed using standard data cleansing and pre-processing approaches to ensure data quality based on the following filtration criteria.

1. Remove twitter handles “@” in order to get only the twitterers’ usernames.

2. Remove the following in order to get only content: Twitter hashtags ”#”, URLs and hyperlinks, Punctuations, and Emoji.

3. Correct and unify the encoding format as some tweets include some complex characters such as â, €™, œ, ¦, â€, etc. thus all tweets are decoded with UTF-8 standard format to transform such symbols to understandable data output.

There are several comprehensive metrics used in pre-processing twitter data particularly for sentiment analysis such as handling negation and removal of duplicate tweets (Arias, Arratia, & Xuriguera, 2014). These metrics are important for sentiment classification. However we only consider semantics of Twitter textual data to define its domain hence those metrics are not necessary for this study. Negation and duplication of retweets are then not considered in this work. Internet slang e.g. ‘lol’ or acronyms and typos are collected and processed within the text mining approach. The ‘lol’ slang is not relevant to the task we are trying to accomplish however some acronyms and typos can be relevant and may not be detected, an area which may be alleviated in future work.

### 3.2 Domain Knowledge Inference

In the domain knowledge inference process, the domain knowledge being captured in domain ontologies is identified and used in the enrichment of the semantics of the tweets. In each tweet that users post, the semantics and the domains of the tweets can be extracted; the extracted domain knowledge is then used to enrich the tweets.

The inference process consists of two stages i.e. start-up stage and learning stage. The start-up stage is a set-up stage that uses AlchemyAPI to identify domain ontologies. Figure 2 shows the domain knowledge inference process during the start-up stage.

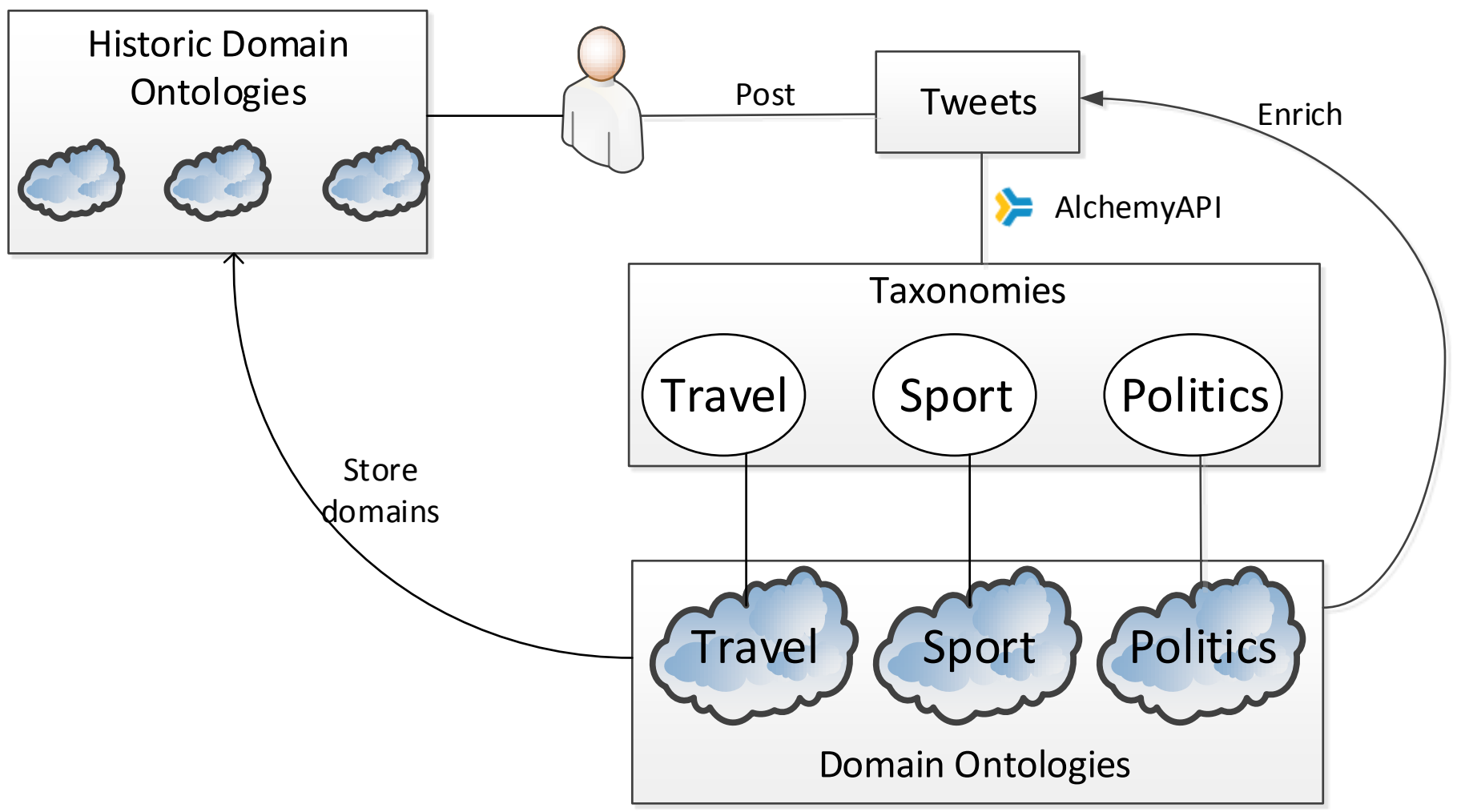

Figure 2: Domain knowledge inference process during the start-up

As shown in the figure, the process starts when a user tweets. Each textual tweet data is processed by AlchemyAPI to obtain its taxonomy. AlchemyAPI provides three domains as taxonomies. Each domain ontology is identified based on the taxonomy and is then used in the enrichment process. For example, AlchemyAPI identifies three taxonomies for a tweet i.e. Travel, Sport, and Politics so three domain ontologies of Travel, Sport, and Politics are assigned as domain knowledge. The three ontologies will be used in the enrichment process and are stored as historic domain ontologies for the particular user who posted the particular tweet.

Once users have a list of historic domains, it is possible to move into the learning stage where machine learning is utilised. Machine learning ranks the historic domains and based on the ranking it provides ability to select the particular domain ontologies for the enrichment process. Domain ontologies are ranked in an orderly number of tweets posted most. Rule based learning is applied here. For example if a user has posted the most tweets about sport, the sport domain ontology is firstly used for tweet enrichment. In a case of user being a celebrity, domain(s) will be obvious and the particular domain ontologies can be applied in the enrichment process. Figure 3 shows the domain knowledge inference process during the learning stage.

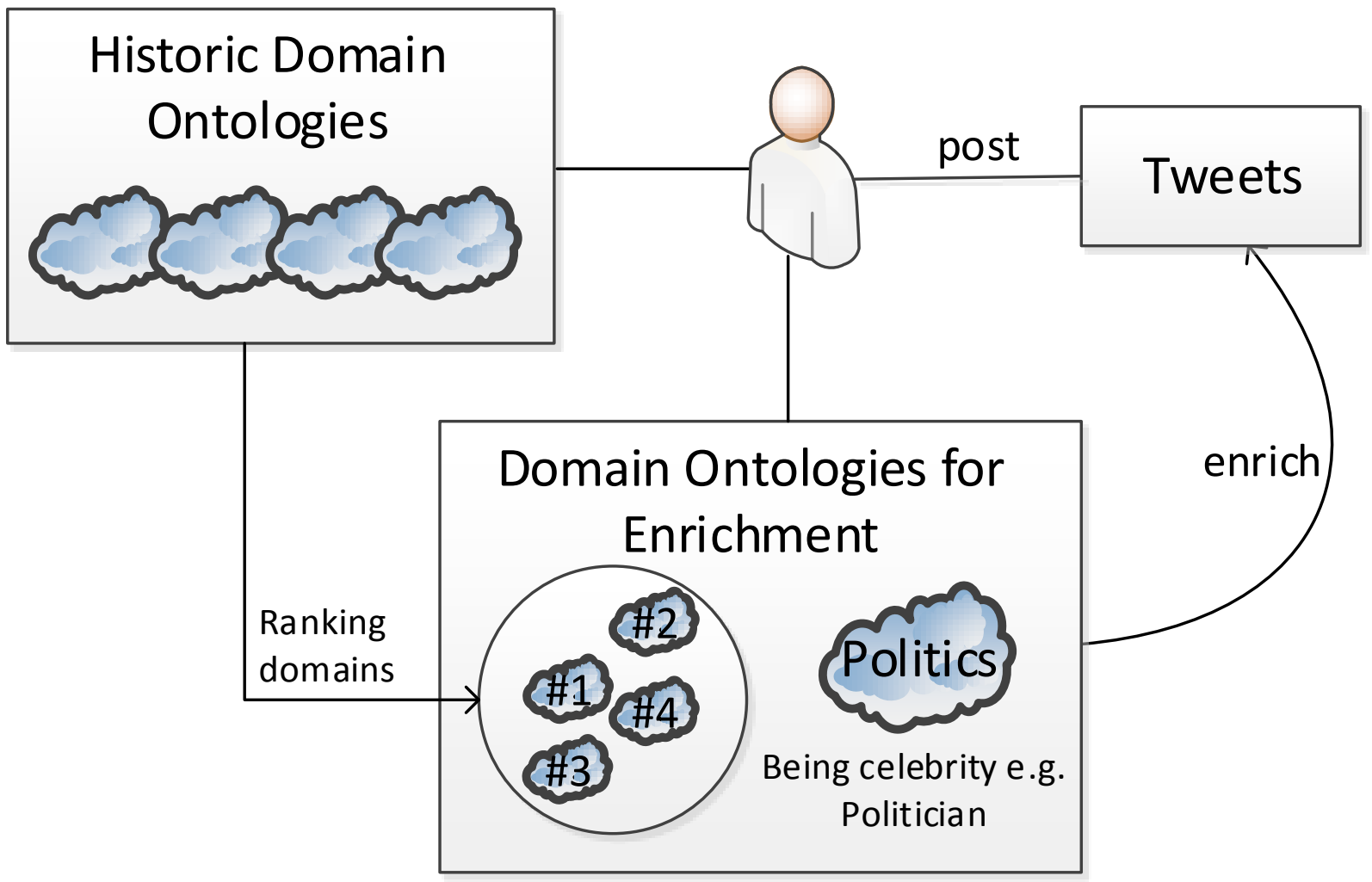

Figure 3: Domain knowledge inference process during learning stage

### 3.3 Annotation and Enrichment

For the annotation and enrichment process, the textual data of tweets is semantically annotated with the concepts in the domain ontologies; the annotation is then enriched with a description of the concepts referring to the domain ontologies and using controlled vocabularies e.g. Dublin Core (DC[4]), Simple Knowledge Organization System (SKOS[5]), Semantically-Interlinked Online Communities (SIOC[6]). This allows each entity in the textual data to be specified with its semantic concept. The particular concepts can be further expanded into other related concepts and other entities instantiated by the concepts. The consolidation of this semantic information provides a detailed view of the entity captured in domain ontologies. We manipulate the domain ontologies using Apache Jena API. Jena, which is a Java framework for building semantic web applications, provides functionalities of create, read, and modify triples (subject – predicate – object) in ontologies.

### 3.4 Interlinking

[4] dublincore.org/
[5] http://www.w3.org/2004/02/skos/
[6] http://sioc-project.org/

For the interlinking process, entities are interlinked with similar entities defined in other datasets to provide an extended view of the entities represented by the concepts. Our focus is on equivalence links specifying URIs (Universal Resource Identifiers) that refer to the same resource or entity. Ontology Web Language (OWL) provides support for equivalence links between ontology components and data. The resources and entities are linked through the 'owl#sameAs' relation; this implies that the subject URI and object URI resources are the same. Hence, the data can be explored in further detail. In the interlinking process, different vocabularies i.e. Upper Mapping and Binding Exchange Layer (UMBEL[7]), Freebase[8] – a community-curated database of well-known people, places, and things, YAGO[9] – a high quality knowledge base, Friend-of-a-Friend (FOAF[10]), Dublin Core (DC[11]), Simple Knowledge Organization System (SKOS[12]), Semantically-Interlinked Online Communities (SIOC[13]), and DBPedia[14] knowledge base, are used to link and enrich the semantic description of resources annotated.

### 3.5 Semantic Repository

A semantic repository represents a knowledge base which continues and updates the semantically rich annotated structured data. Ontology formalises the conceptualised knowledge in a particular domain and provides explicit semantics by splitting concepts, their attributes, and their relationships from the instances. In the repository, there are terminological data that define concepts (classes), attributes (data properties), relationships (object properties), and axioms (constraints) as well as data that enumerates the instances (individuals). This enables different services support such as concept-based search, entailment to retrieve implied knowledge, instance-related information retrieval, etc. By using the semantic repository, we can perform query expansion for entity disambiguation and to retrieve semantic description of entities.

[7] http://umbel.org/
[8] http://www.freebase.com/
[9] http://www.foaf-project.org/
[10] http://www.foaf-project.org/
[11] dublincore.org/
[12] http://www.w3.org/2004/02/skos/
[13] http://sioc-project.org/
[14] wiki.dbpedia.org/

In the repository, the structured data are stored as the RDF graph for persistence. Virtuoso (open source edition) triple store is used to store the RDF triples, ontologies, schemas, and expose it using a SPARQL endpoint. The SPARQL endpoint enables applications, users, software agents, and the like to access the knowledge base by posing SPARQL queries.

## 4. System Components

### 4.1 Politics Ontology

The BBC[15] produces a plethora of rich and diverse content about things that matter to the BBC's audiences ranging from athletes, politicians, or artists ("BBC Ontologies," 2015). The BBC uses domain Ontologies to describe the world and content the BBC creates and to manage and share data within the Linked Data platform. Linked Data provides an opportunity to connect the content together through various topics. Among the nine domain ontologies that the BBC has developed and uses, the Politics Ontology describes a model for politics, specifically in terms of local government and elections (Berlanga et al., 2014). This was originally designed to cope with UK (England and Northern Ireland) Local, and European Elections in May 2014. The focus of the project is on Australian Politics hence we have developed a domain-specific Politics Ontology for Australian Politics by extending the BBC Politics Ontology. We specified the ontology in Australian Politics having Australian politicians and Australian political parties. The concepts, instances, and relations are used in the annotation process. At this stage, the concept Politician has 53 instances of Australian politicians and the concept Political Party has 4 instances of Australian political parties. The politics ontology is being incrementally extended over time. Figure 4 shows the BBC Politics Ontology; Figure 5 shows the extended version of the BBC Politics Ontology using OntoGraf for visualisation of the relationships in ontologies.

---

[15] http://www.bbc.com/

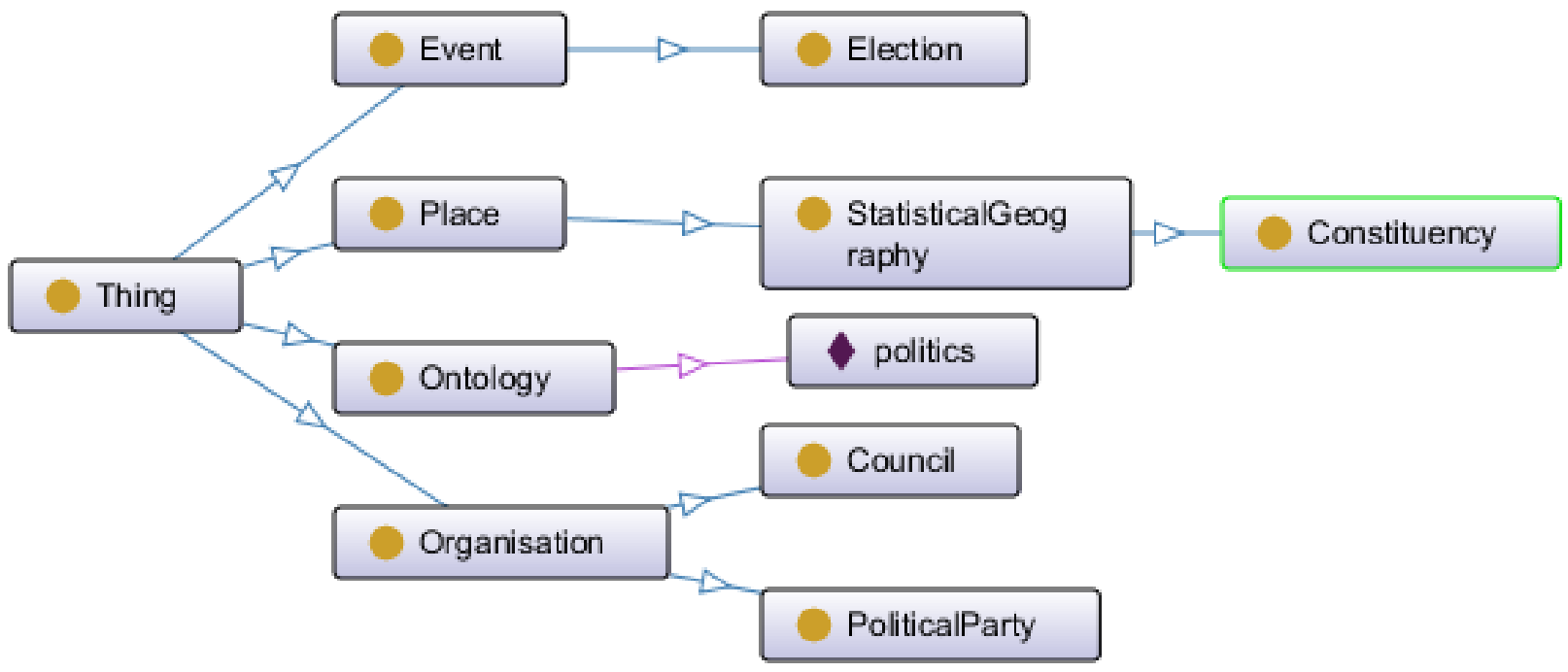

Figure 4: BBC politics ontology

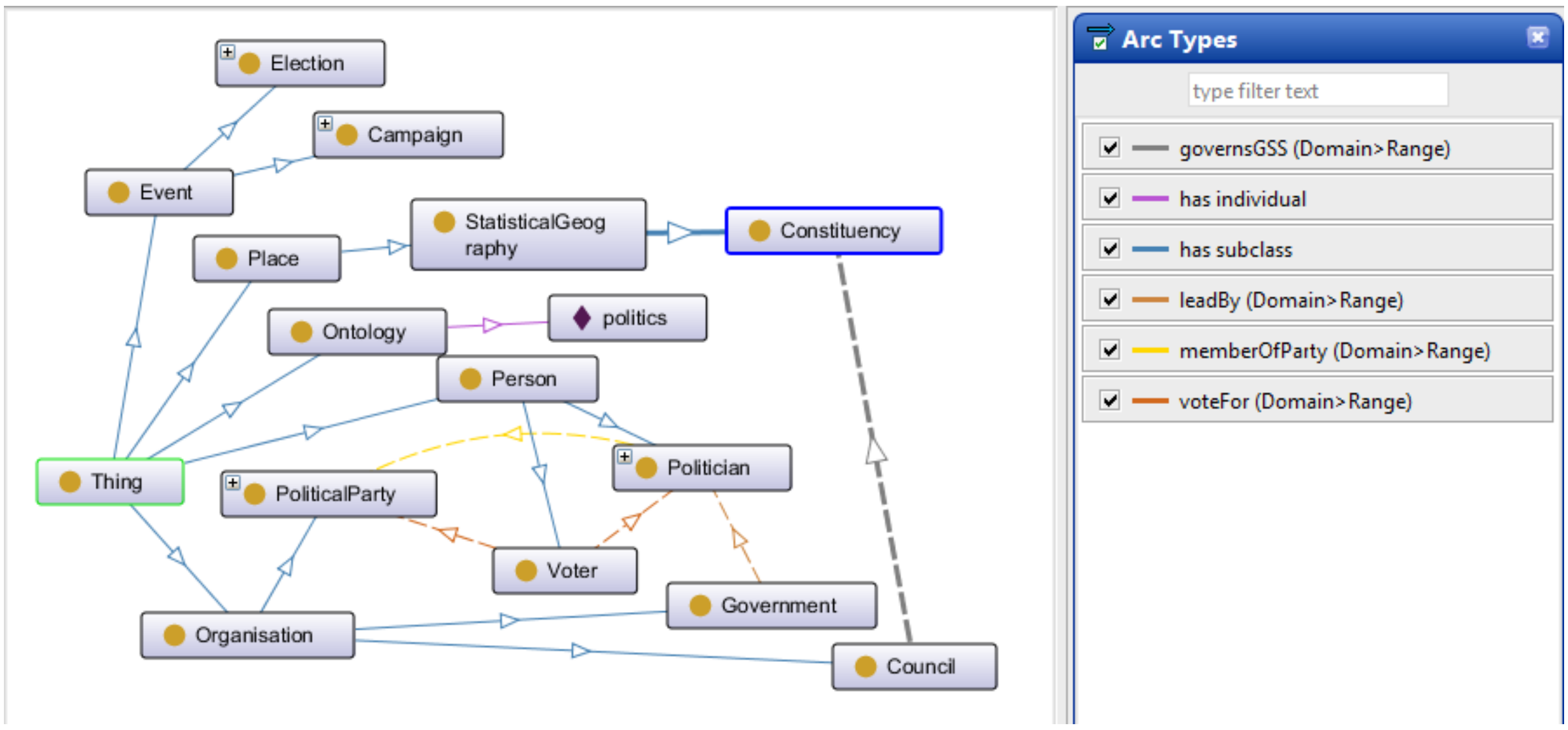

Figure 5: BBC politics ontology extension

In order to ensure the extended version of Politics Ontology is consistent, which is important as part of an ontology's development and testing, the Ontology needs to undergo a reasoning process. No reliable conclusion can be deduced otherwise. The extended version of the Politics Ontology has been reasoned to check its logical consistency using FaCT++, HermiT, Pellet, Pellet (Incremental), RacerPro and TrOWL reasoners. The reasoners checked the class, object/data property hierarchies, the class/object property assertions, and whether there were the same individuals contained within the ontology. Consistency verification through a reasoner includes consistency checking, concept satisfiability, classification, and realisation which are all standard inference services conventionally provided by a reasoner. The extended version of the Politics Ontology does not contain any contradictory facts.

### 4.2 Text Mining Tool

Text mining techniques have been applied for entity recognition, text classification, terminology extraction, and relationship extraction (Cohen & Hersh, 2005). In order to convert unstructured textual data from large scale collections to a semi-structured or structured data filtering based on the need, natural language processing algorithms are used (Bello-Orgaz et al., 2016). However this can be difficult because the same word can mean different things depending on context. Ontologies can help to automate human understanding of the concepts and the relationships between concepts. Ontologies allow for achieving a certain level of filtering accuracy. Hence in this paper we use text mining tool together with domain specific ontologies for better accuracy of concept identification.

There are several text mining tools that can extract entities and map the entities with concepts for online textual data. Rizzo and Troncy (Rizzo & Troncy, 2011) evaluate five popular entity extraction tools on a dataset of news articles i.e. AlchemyAPI, Zemanta, OpenCalais, DBPedia Spotlight, and Extractiv. Saif et al. (Saif et al., 2012) chose to evaluate the first three of the five entity extraction tools on tweets. The results from experiments in both studies consistently show that AlchemyAPI performed best for entity extraction and semantic concept mapping. In addition, in March 2015 IBM has acquired Alchemy for development of next generation cognitive computing applications offered under IBM's Watson ("IBM Acquires AlchemyAPI, Enhancing Watson's Deep Learning Capabilities," 2015). Hence, we use and evaluate the use of Alchemy in our project. Evaluation of other tools can be done in future work.

AlchemyAPI uses natural language processing, machine learning algorithms and deep learning models to power its core technology ("What are the algorithms behind AlchemyAPI? - Quora,"). It's not an open-source technology hence the algorithm, statistical method, and mathematical method used in Alchemy are not released ("nlp - Algorithms behind the Alchemy API for concept and keywords extraction - Stack Overflow,").

**4.3 Politic twitter data**

We used REST API to collect public archived tweets. For the work and experiments, we run the collected tweet data through AlchemyAPI and select tweets for our dataset based on the set thresholds as follows which are defined by AlchemyAPI:

1. Having confidence score above 0.4 AND
2. Not having confidence response data status as no (not confidence).

AlchemyAPI provides a confidence score for the detected category ranged from 0.0 to 1.0 where higher is better (Turian, 2013). The confidence score and response data conveys the likelihood of the identified category being correct.

Table 2 shows dataset sources, the number of collected tweets and number of selected tweets, and period of collection. The number of collected tweets are those tweets we collected during a period of time however we only select number of tweets for experiments based on thresholds above mentioned. In order to get politics data, politicians are the main source and we consider journalists tweets as an addition source. The two datasets contain politics data; the difference is that one from politicians' view and the other from journalists' view. Both datasets are chosen for experiments of this research.

Table 2: Details of two datasets

| | Sources | No. of collected tweets | No. of selected tweets | Period of collection |
|---|---|---|---|---|
| Politics dataset | Twitter accounts of two Australian politicians. | 4,122 (1,954 and 2,168) | 3,653 | 25th Jan 2011 - 26th March 2015 |
| Politics Influence dataset | Twitter accounts of two Australian journalists. | 3,479 (3,207 and 272) | 210 | 5th Oct 2010 - 20th May 2015 |

## 5. Case Studies of Politics Twitter Data

AlchemyAPI is able to identify people, companies, organizations, cities, geographic features, and other types of entities from the textual data content in the general classification. It supports Linked Data and employs natural language processing technology to analyse the data and extract the semantic richness embedded within (Turian, 2013). It is a comprehensive tool however it can only categorize the most general classification due to the lack of domain specific knowledge. For specific domain, AlchemyAPI will need ontologies to categorise content based on ontology concepts, instances, and relationships. Hence, the ontology based approach proposed in this paper will be of benefit in terms of extending the existing AlchemyAPI.

An example of output from AlchemyAPI for entity extraction, concepts mapping, and taxonomy classification of a tweet is shown in Figure 6.

Tweet: "Launched Jennifer Kanis for Melbourne Campaign today. Outcomes instead of ineffective self indulgent commentary. Vote Labor in Melbourne."

AlchemyAPI entity extraction and concept mapping results:

ENTITY: Jennifer Kanis; TYPE of ENTITY: Person

ENTITY: Melbourne Campaign; TYPE of ENTITY: Organization

ENTITY: Melbourne; TYPE of ENTITY: City

AlchemyAPI taxonomy results:

/travel/tourist destinations/australia and new zealand

/society/work/unions

Figure 6: Output from AlchemyAPI for entity extraction, concepts mapping, and taxonomy classification of a tweet

As can be seen, Alchemy fails to capture the keywords 'vote' and 'labor' as entities due to lack of specific domain knowledge. As result, the taxonomy classifications of travel and society are inadequate. However, if politics ontology is applied as specific domain knowledge, the keywords 'Vote' and 'Labor' are annotated with its type respectively as relation 'voteFor' and concept 'Political Party'. By annotating two more entities of Labor and Vote and specifying particular entity Jennifer Kanis as Politician as shown in Figure 7, the politics domain is counted as domain of this tweet in addition to the travel and society domains. The more data that are annotated, the more entities are extracted in which the domain of tweet is clearer.

Jennifer Kanis - CONCEPT: Politician

Labor - CONCEPT: Political Party

Vote – Relation: voteFor

Figure 7: Politics Ontology Annotation

In addition, based on the credibility domain of politics, the entities can be inferred to the knowledge captured in the Politics ontology. Figure 8 shows entities 'Jennifer Kanis', 'Labor', and 'Vote' being respectively inferred to concepts 'Politician' and 'Political Party' and relation 'voteFor'. As can be seen in Figure 8, the concept 'Politician' relates to the concept 'Person' and the concept 'Political

Party' relates to the concept 'Organisation' through a generalisation relationship. The concept 'Political Party' relates to the concept 'Politician' through the associated relationship 'memberOf' which is the converse of the associated relationship 'ledBy'. This forms as the domain of knowledge in politics. Table 3 shows the modelling notations that appear in Figure 8.

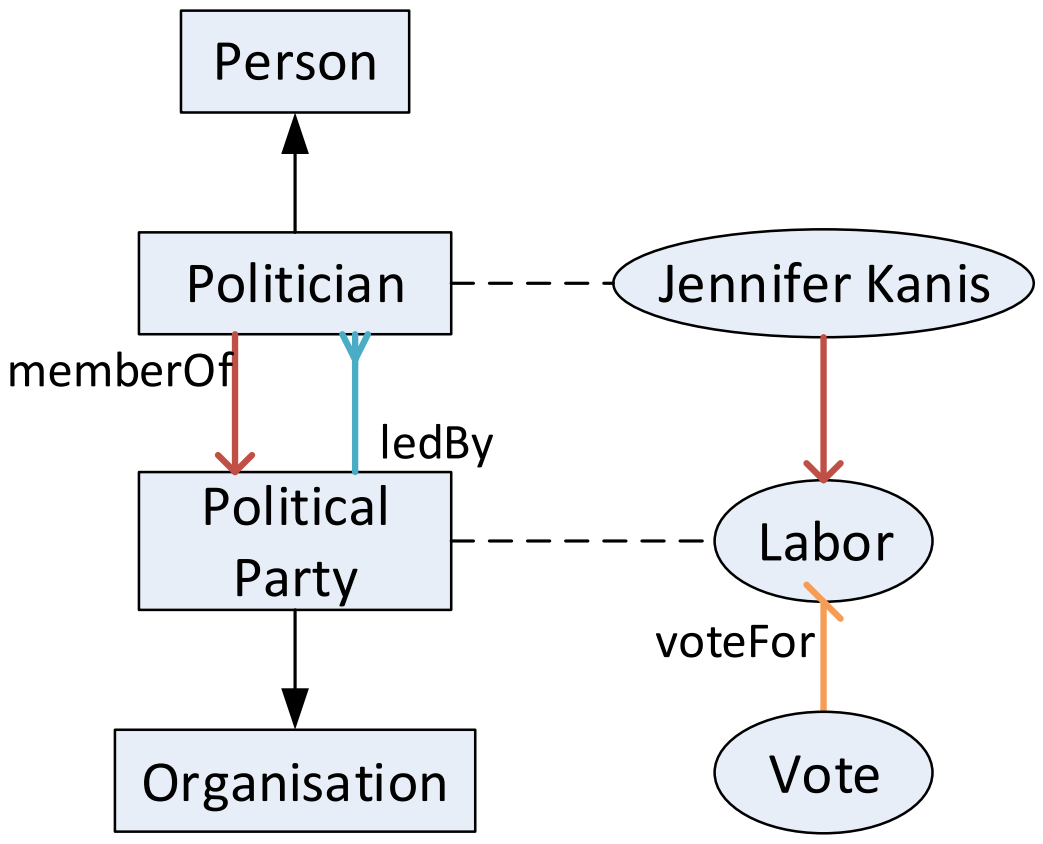

Figure 8: Knowledge captured in politics ontology

Table 3: Ontology modelling notations

| Notations | Semantics |
|---|---|
| 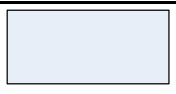 | Concept / Ontology class |
| 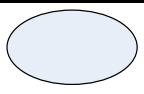 | Instance / Individual |
| ⟶ | Association semantical relationship (different colours and different end arrow types represent different relationships) |
| ➞ | Generalisation / Taxonomical / Hierarchical relationship |
| _ _ - | Instance / Individual relationship |

Hence, by integrating the results of the AlchemyAPI and the politics ontology annotation, we can infer the following information from the particular tweet:

1. Jennifer Kanis is a Politician; Politician is a Person.

2. Labor is a Political Party; Political Party is an Organisation.

3. Jennifer Kanis is a member of Labor.

4. Vote for Labor.

5. Melbourne is a city.

Figure 9 indicates the query and subsequent result to retrieve all information of Labour party. As can be seen, it shows entity 'Labour' enriched with its type of political party, website, and official name. The entity can also interlink with controlled vocabularies. Here, the entity 'Labour' is interlinked with vocabularies from dbpedia, freebase, yago, and semanticweb.

**Query**

```
PREFIX Politics: <http://www.semanticweb.org/ontologies/Politics.owl#>
SELECT *
        WHERE { Politics: labour ?b ?c}
```

**Result**

| b | c |
|---|---|
| http://www.w3.org/1999/02/22-rdf-syntax-ns#type | http://www.semanticweb.org/owl/owlapi/turtle#PoliticalParty |
| http://www.w3.org/2002/07/owl#sameAs | http://dbpedia.org/resource/Australian_Labor_Party |
| http://www.w3.org/2002/07/owl#sameAs | http://rdf.freebase.com/ns/m.0q96 |
| http://www.w3.org/2002/07/owl#sameAs | http://yago-knowledge.org/resource/Australian_Labor_Party |
| http://www.w3.org/2002/07/owl#sameAs | http://www.semanticweb.org/owl/owlapi/turtle#Labor |
| http://www.semanticweb.org/owl/owlapi/turtle#ResolvedName | "Australian Labor Party" |
| http://www.semanticweb.org/owl/owlapi/turtle#Website | "http://www.alp.org.au/" |
| http://www.semanticweb.org/owl/owlapi/turtle#value | "labour" |

Figure 9: Enrichment and interlinking of Labour party

Figure 10 provides the query that retrieves all information of Politician Daniel Andrews. As can be seen, it shows the enrichment and interlinking of the entity with its name, its type of Politician, and its subclass of Person. The entity is also interlinked with vocabularies from dbpedia, freebase, yago, and semanticweb.

## Query

```
PREFIX Politics: <http://www.semanticweb.org/ontologies/Politics.owl#>
SELECT *
        WHERE { Politics: DanielAndrews ?p ?o}
```

## Result

| p | o |
|---|---|
| http://www.w3.org/1999/02/22-rdf-syntax-ns#type | http://www.semanticweb.org/owl/owlapi/turtle#Politician |
| http://www.w3.org/2000/01/rdf-schema#subClassOf | http://www.semanticweb.org/owl/owlapi/turtle#Person |
| http://www.w3.org/2002/07/owl#sameAs | http://dbpedia.org/resource/Daniel_Andrews |
| http://www.w3.org/2002/07/owl#sameAs | http://rdf.freebase.com/ns/m.0bwttx |
| http://www.w3.org/2002/07/owl#sameAs | http://yago-knowledge.org/resource/Daniel_Andrews |
| http://www.w3.org/2002/07/owl#sameAs | http://www.semanticweb.org/owl/owlapi/turtle#DanielAndrews |
| http://www.semanticweb.org/owl/owlapi/turtle#ResolvedName | "Daniel Andrews" |
| http://www.semanticweb.org/owl/owlapi/turtle#value | "danielandrewsmp" |

Figure 10: Enrichment and interlinking of Politician Daniel Andrews

Another example is shown in Figure 11.

Tweet: "Thoughts and prayers with Karen Overington's family today. Karen was true Labor, a true friend and will be truly missed by all of us."

AlchemyAPI entity extraction and concept mapping results:

ENTITY: Karen Overington; TYPE of ENTITY: Politician

AlchemyAPI taxonomy results:

/society/work/unions

/family and parenting

Figure 11: Output from AlchemyAPI for entity extraction, concepts mapping, and taxonomy classification of a tweet

In the tweet shown in Figure 11 above, AlchemyAPI captures only the entity ‘Karen’ Overington as politician. The entity and keywords of ‘true friends’, ‘prayers’, ‘thoughts’, ‘family’, and ‘labor’ are used to classify the tweet under the taxonomy of society and family and parenting which is inadequate. Hence, if politics ontology is applied, the keyword ‘labor’ is annotated as an entity under the concept of political party. This results in classifying the Politics domain as an additional domain of tweet.

We have experimented with the politics dataset. AlchemyAPI classifies the politics dataset into various domains as shown in Figure 12. For two different users, it shows that most tweets are in the travel domain though it is supposed to be in politics domain due to the politics dataset. In comparison to results from AlchemyAPI associated with the Politics ontology as shown in Figure 13, it classifies the same dataset into the proper domain i.e. the politics domain. This shows significant improvement when associated with specific domain knowledge of politics being captured in the Politics ontology.

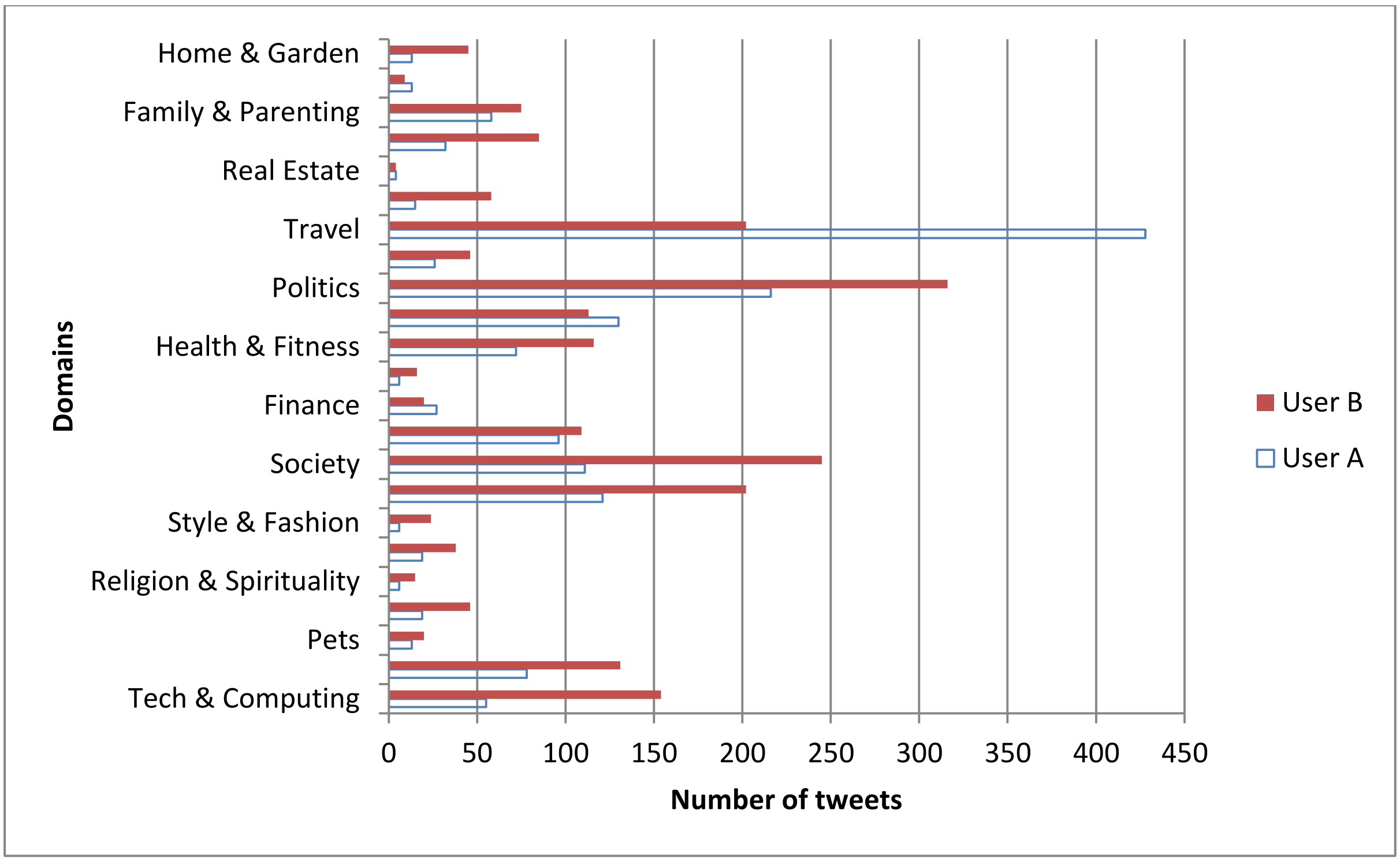

Figure 12: Results from Alchemy showing a number of tweets in various domains from the politics dataset

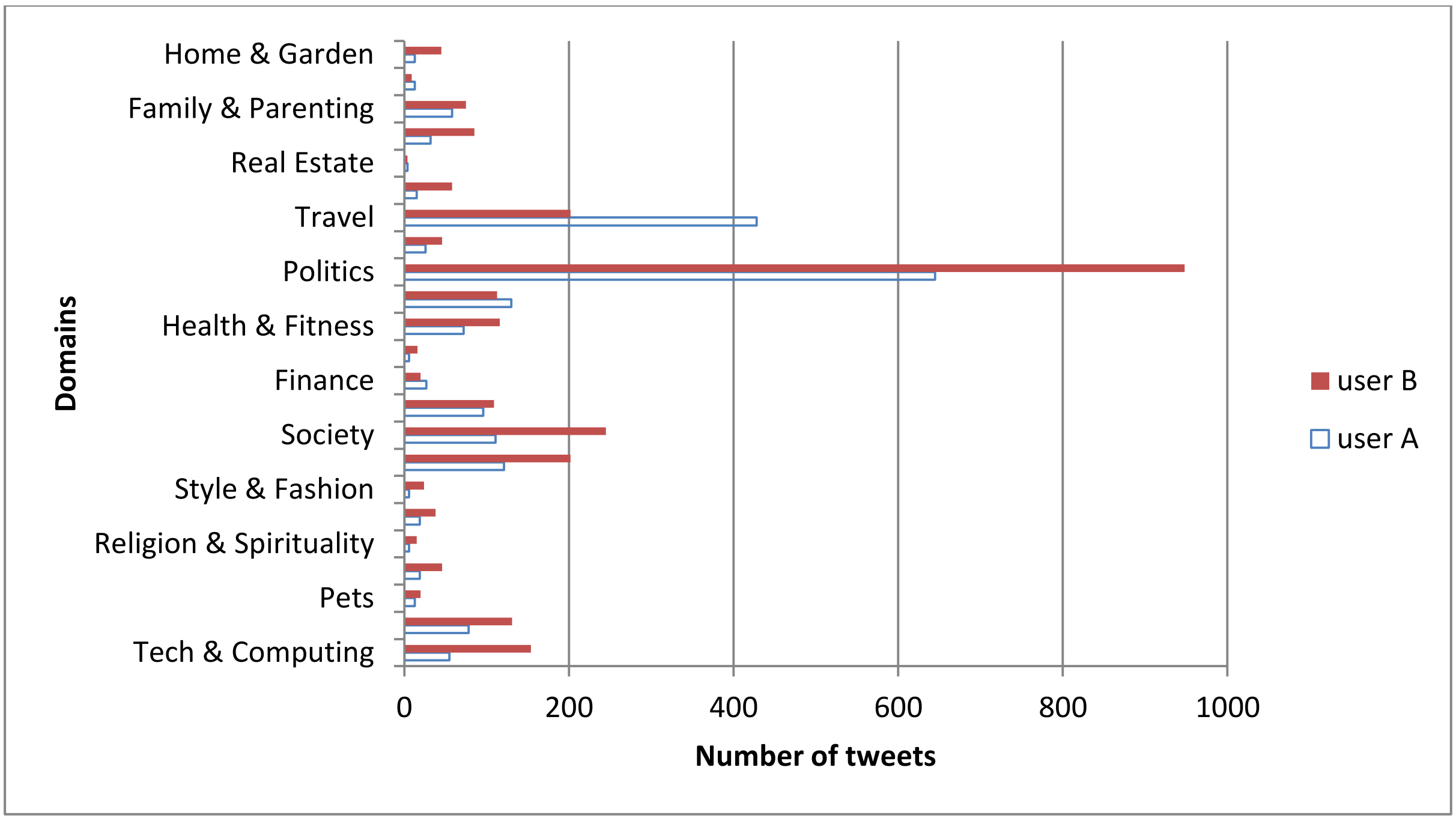

Figure 13: Results from Alchemy associated with politics ontology showing a number of tweets in various domains from the politics dataset

Once the domain can be correctly defined from user's tweets using the proposed ontology based approach, users' influence in particular domains can be discovered and domain-based trustworthiness can also be evaluated. Domain-based trustworthiness evaluation is discussed in section 7 which covers source reputation and user's trustworthiness. In next section, we evaluate our approach for domain classification and entity annotation.

## 6. Evaluation

In this section, we evaluate the semantic information extraction at the domain level and the entity level. We compare the performance of AlchemyAPI alone with the performance of AlchemyAPI when it is associated with our ontology-based approach.

### 6.1 Datasets

For evaluation purpose, we chosen 473 tweets from the selected politics dataset and chosen 209 tweets from the selected politics-influenced dataset. We divide datasets for evaluation into 4 categories:

1. Category #1: Tweets that are classified by AlchemyAPI as in the politics domain and the Politics ontology annotates them.
2. Category #2: Tweets that are classified by AlchemyAPI as in the NON politics domain however the Politics ontology annotates them.
3. Category #3: Tweets that are classified by AlchemyAPI as in the politics domain but the Politics ontology does NOT annotate them.
4. Category #4: Tweets that are classified by AlchemyAPI as in the NON politics domain and the Politics ontology does NOT annotate them.

### 6.2 Evaluators

Three evaluators are used to evaluate the concept extraction and domain identification outputs generated by AlchemyAPI alone compared with AlchemyAPI associated with our ontology-based approach. One of the evaluators is considered as a domain expert in politics i.e. this person is currently involved in politics and has worked in the area for more than five years. The other two evaluators are academics and considered non-domain experts who have a general interest in the politics domain.

### 6.3 Results and Discussion

The assessment of the outputs is based on

1. the correctness of the extracted politics entities;
2. the correctness of inferring the extracted politics entities with its concept; and
3. the correctness of politics domain classified in tweets.

#### 6.3.1 Politics dataset

This section discusses the evaluation results from the politics dataset. Table 4 shows the number of *correct* extracted politics entities. The results show that for tweets that are classified by AlchemyAPI as politics tweets, the politics ontology can annotate 98 more politics entities from just 41 entities from AlchemyAPI . The number of politics entities increases to 139 entities when combining the AlchemyAPI result with the politics ontology result; that is, the number of entities is almost tripled. For the non-politics tweets classified by AlchemyAPI, the politics ontology can annotate 161 more politics entities from just 62 entities from AlchemyAPI. The number of politics entities increases to 223 entities when combining the AlchemyAPI result with the politics ontology result, i.e. almost 4 times more entities.

Table 4: Number of *correct* extracted politics entities

| Categories of dataset | AlchemyAPI | Politics Ontology | AlchemyAPI and Politics Ontology |
|---|---|---|---|
| Alchemy Politics tweet being annotated by Politics ontology | 41 | 98 | 139 |
| Alchemy NON-Politics tweet being annotated by Politics ontology | 62 | 161 | 223 |
| Alchemy Politics tweet NOT being annotated by Politics ontology | 0 | 0 | 0 |
| Alchemy NON-Politics tweet NOT being annotated by Politics ontology | 0 | 0 | 0 |
| Total | 103 | 259 | 362 |
| Percentage of *correct* extracted entities (sample size of 473) | 22% | 55% | 77% |

Table 5 shows the number of *incorrect* extracted politics entities in the 4 categories as explained in Section 6.1 for datasets. The results show some flaws in AlchemyAPI which can be overcome by incorporating it with specific domain knowledge captured in politics ontology.

Table 5: Number of *incorrect* extracted politics entities

| Categories of dataset | AlchemyAPI |
|---|---|
| category #1: Alchemy Politics tweet being annotated by Politics ontology | 35 |
| category #2: Alchemy NON-Politics tweet being annotated by Politics ontology | 35 |
| category #3: Alchemy Politics tweet NOT being annotated by Politics ontology | 8 |
| category #4: Alchemy NON-Politics tweet NOT being annotated by Politics ontology | 8 |
| Total | 86 |

In total, AlchemyAPI alone extracts 103 politics entities, failing to extract 259 politics entities which the politics ontology annotates as entities. Hence, by incorporating the politics ontology with AlchemyAPI, more politics entities are extracted, totalling 362 entities rather than just 103 entities.

The pie chart shows all distinct 681 entities resulting from AlchemyAPI as seen in Figure 14. The results show that AlchemyAPI identifies more entities in other domains outside the politics domain in the politics dataset.

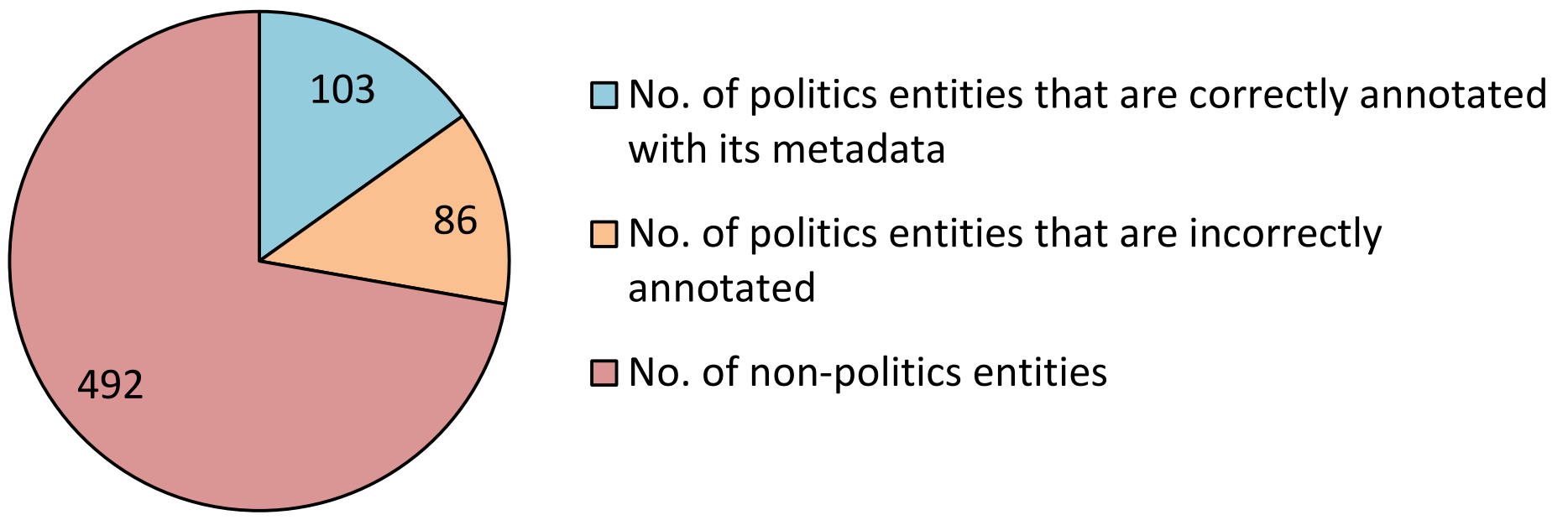

Figure 14: All distinct entities resulting from AlchemyAPI

### 6.3.2 Politics-influenced dataset

This section discusses evaluation results from politics influence dataset. Table 6 shows that AlchemyAPI alone correctly extracts 44 politics entities, incorrectly extracts 15 politics entities, and fails to extract 59 politics entities which the politics ontology annotates as entities. By incorporating the politics ontology with AlchemyAPI, more politics entities are extracted, totalling 103 entities which is over twice the number of entities extracted by AlchemyAPI alone.

Table 6: Politics entity extraction in AlchemyAPI from politics-influenced dataset

| | Correct | Incorrect | Missing politics | Total number of | Total number of |
|---|---|---|---|---|---|

| | extracted politics entities | extracted politics entities | entities | retrieved entities | politics entities |
|---|---|---|---|---|---|
| AlchemyAPI | 44 | 15 | 59 | 59 | 103 |

### 6.3.3 Precision, recall, and F-measure

In this section, we show precision, recall, and F-measure from AlchemyAPI results for both datasets. Precision is the fraction of retrieved entities that are politics-related as shown in equation (1) while recall is the fraction of politics entities that are retrieved as shown in equation (2). Another metric known as the F-measure, which is the weighted harmonic mean of precision and recall, is used as shown in equation (3).

Precision = Number of Politics Entities Retrieved / Total Number of Retrieved Entities (1)

Recall= Number of Politics Entities Retrieved / Total Number of Politics Entities (2)

$$\text{F-measure} = 2 \times \frac{precision \times recall}{precision + recall} \quad (3)$$

Figure 15 shows a comparison of politics data and politics-influenced data on precision, recall, and F-measure. From the figure, it can be observed that AlchemyAPI performs better in data from various domains (politics influence dataset) rather than domain-specific data (politics dataset).

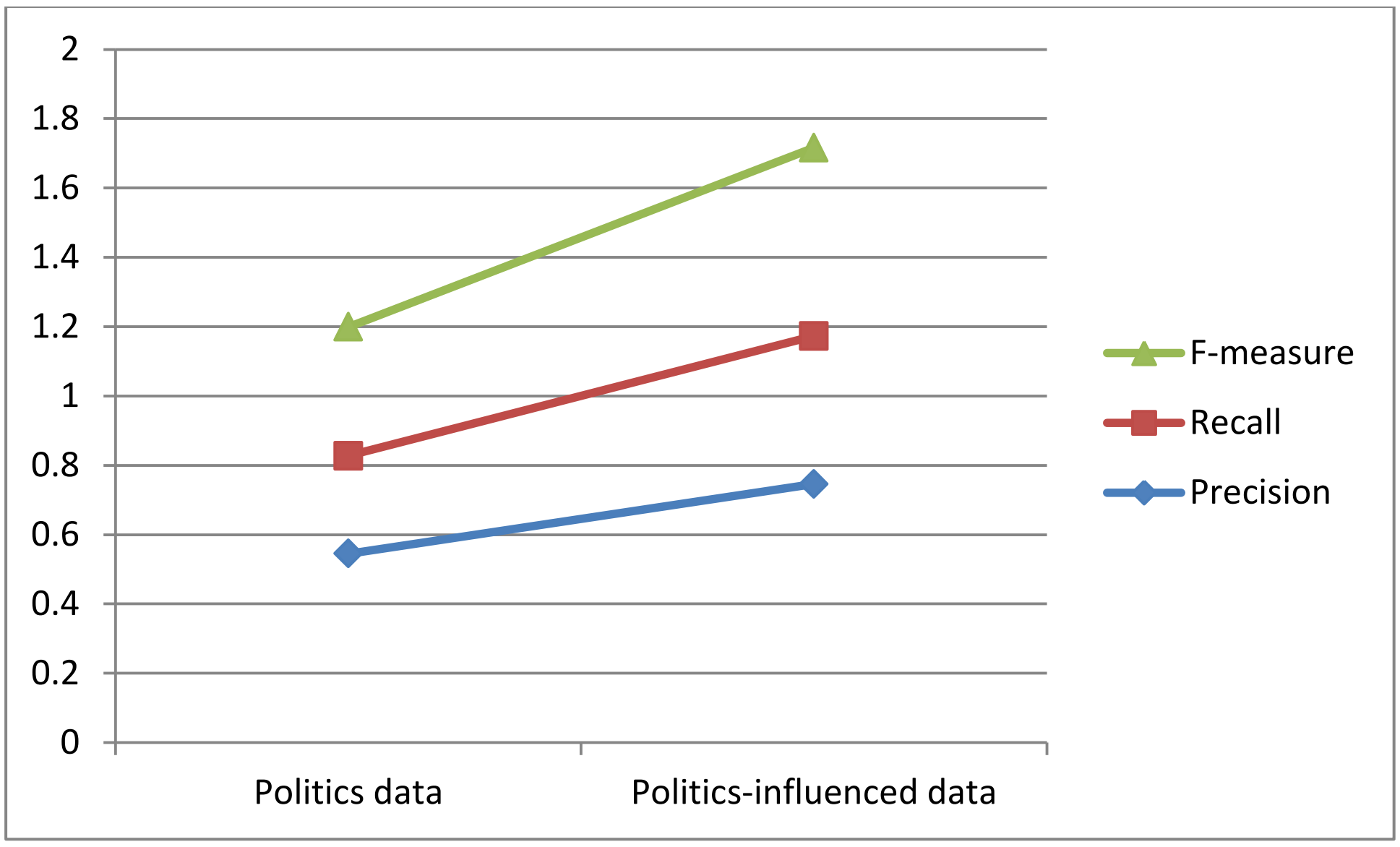

Figure 15: Comparison of politics data and politics influence data on precision, recall, and F-measure

In terms of precision, it shows that AlchemyAPI can retrieve more entities that are politics-related in a politics-influenced dataset than in a politics dataset. This is because fewer incorrect politics entities are retrieved from a politics-influenced dataset.

In terms of recall, it shows that AlchemyAPI can retrieve more politics entities in a politics-influenced dataset than in a politics dataset. This is because AlchemyAPI should have identified more politics entities in the politics dataset, but failed to do so.

### 6.3.4 Politics domain classification

In this section, we show the correctness of the politics domain classified in tweets. The evaluators validate each tweet in the datasets and determine whether it is a politics-related post. Table 7 shows the percentage of tweets being classified as politics-related.

Table 7: Percentage of tweets being classified as politics-related

| Categories of dataset | Politics dataset | Politics-influenced dataset |
|---|---|---|
| Alchemy Politics tweet being annotated by Politics ontology | 99% | 98% |
| Alchemy NON-Politics tweet being annotated by Politics ontology | 98% | 97% |
| Alchemy Politics tweet NOT being annotated by Politics ontology | 27% | 47% |
| Alchemy NON-Politics tweet NOT being annotated by Politics ontology | 12% | 32% |

It can be observed from the results that almost all tweets that the politics ontology annotates are politics tweets. The politics ontology annotates less than 50% of the politics tweets, but more of the politics-influenced dataset than the politics dataset. This indicates that domain-specific ontology performs better in a domain-specific dataset.

## 7 Future Research Directions in Domain-based Trustworthiness

As mentioned, this manuscript reports on work in progress. In the next stage of the project, we intend to apply the ontology-based approach in social business intelligence (i) to ascertain the credible information, (ii) to determine the reputation of the sources, and (iii) to define the legitimate contributors with a degree of trustworthy of the information, the sources, and the users.

### 7.1 Social Business Intelligence

In a competitive environment, one of the main challenges over the past few years for an organisation is to understand data and discover its hidden value in order to deliver timely, accurate, and advanced information and knowledge for decision making. The data exists in different types, ranging from structured data in relational databases to unstructured data in file systems and to semi-structured data neither in raw nor strictly typed as in conventional database systems. Structured data is usually produced by the day-to-day operational activities of a business. However, most of the businesses also produce unstructured or semi-structured data that need to be discovered i.e. those data produced by communication between business and customer such as customer feedback, contracts, complaint emails or transcripts of telephone conversations. Moreover, the widespread increase of social media such as Facebook, Twitter, Instagram, Flickr and YouTube has provided opportunities to businesses to study customer views and market data at very large scales and for very large populations (De Choudhury et al., 2010). As a result, analysts today are able to conduct in-depth analysis of external business data such as customer blog postings (Gruhl, Guha, Liben-Nowell, & Tomkins, 2004), Internet chain-letter data (Liben-Nowell & Kleiberg, 2008), social tagging (Anagnostopoulos, Kumar, & Mahdian, 2008), Facebook news feed (Sun, Rosenn, Marlow, & Lento, 2009) and many other data sources. Social business intelligence aims to cover these data formats and collect these data from different data sources such as operational databases, web logs, social media and other useful sources.

### 7.2 Reputation of the sources

Data sources have increased from transactional data sources and limited external data sources to many other data sources such as data coming from global environment in the form of news, economic factors, etc. and from the Voice of the Market and the Voice of the Customer in the form of social networks, web blogs, etc. All external data sources do not have the same reputation. For example, data coming from news agencies or highly trusted web blogs are more valuable than data coming from poorly trusted web blogs or comments posted in social networks. Similarly, all comments posted in social networks do not have the same impact. For example, comments of users who have a high number of followers have more impact than comments from new users or those with a small number of followers. Knowledge that is generated by using highly trusted data sources and/or which has a high impact factor, raises confidence levels when decision-making. Searching the deep web and assessing the trustworthiness of web files has been identified as the next big challenge for information management (Wright, 2008). The source selection depends on "the trustworthiness of the data in the source and trustworthiness is a measure of correctness of the answer. For example, for the query 'The Godfather', many databases in Google return copies of the book with unrealistically low prices to attract the user attention. When the user proceeds towards the checkout, these low priced items would turn out to be either out of stock or a different item with the same title and cover" (p. 227) (Balakrishnan & Kambhampati, 2011). There are several techniques for measuring the trustworthiness of an external data source. One such method is the CCCI (Correlation, Commitment, Clarity, and Influence) method. CCCI determines the correlation between the originally committed services and the services actually delivered by a Trusted Agent in a business interaction over the service-oriented networks in order to determine the trustworthiness of the Trusted Agent (Chang, Hussain, & Dillon, 2005). This method uses a scale as a measurement system to determine the level of trust. The scale system can have either numeric measures or non-numeric measures. The trustworthiness measure determines the amount of trust that the Trusting Agent has in the Trusted Agent. One of the most popular scale systems in this method is a 7-level trustworthiness scale system. This trustworthiness measure helps to rate trust by numerically quantifying the trust values and qualifying the trust levels numerically. This method is used by different websites such as eBay, YouTube and most customer-to-customer buying and selling websites to measure the trust level of buyers and sellers, and helps other members to decide whether or not to enter into a transaction with trusted or trusting agents. Several other techniques such as the use of neural networks can also be applied to measure the level of trust between trusted and trusting agents.

### 7.3 Domain-based user's trustworthiness and credibility of information

It is essential to evaluate users' credibility and extract trustworthy information. In any domain of interest, the concrete knowledge captured in ontology is used for comparison to find a degree of truthfulness in considered spatial and temporal attributes. Most of the existing trustworthiness evaluation approaches of users and their posts in social networks are generic approaches. There is a lack of domain-based trustworthiness evaluation mechanisms. Discovering users' influence in a specific domain has been motivated by its significance in a broad range of applications such as personalized recommendation systems and expertise retrieval. A novel discriminating measurement for users in a set of knowledge domains will be focused. Domains are extracted from the user's textual data posted using the ontology-based approach presented in this paper. In order to ascertain the level of trustworthiness, a metric incorporating a number of attributes extracted from textual data analysis and user analysis will be consolidated and formulated.

It is important to distinguish users in a set of domains. The idea of discrimination was proposed in Information Retrieval (IR) by applying the $\mathrm{tf.idf}$ formula (S. E. Robertson & Jones, 1976). The intuition was that a query term which occurs in many documents is not a good discriminator (S. Robertson, 2004). This implies that a term which occurs in many documents decreases its weight in general as this term does not show the particular document of interest to the user (Ramos, 2003). This heuristic aspect can be incorporated into a model to evaluate the trustworthiness of users. Consequently, we can argue that a user who posts in all domains has a low trustworthiness value in general. This argument is justified based on the following facts: (i) No one person is an expert in all domains (Gentner & Stevens, 1983); (ii) A user who posts in all domains does not declare to other users which domain(s) s/he is interested in. A user shows to other users which domain s/he is interested in by posting a wide range of contents in that particular domain; (iii) There is the possibility that this user is a spammer due to the behaviour of spammers posting tweets about multiple topics (Wang, 2010). This could end up by tweets being posted in all domains which do not reflect a legitimate user's behaviour.

Moreover, a metric incorporating a number of attributes to measure users' behaviours in social networks will be investigated. The key attributes will be obtained from context data analysis and user analysis. A fine-grained trustworthiness analysis of users and their domains of interest can be provided.

# 8 Conclusion and Future Work

The ontology-based approach has been presented as a means of extracting the semantics of textual data. We proposed to capture domain knowledge in ontologies which are then used to enrich the semantics of data with specific semantics conceptual representation of entities. Five steps in the process were presented: pre-processing, domain knowledge inference, annotation and enrichment, interlinking, and semantic repository. We conducted experiments in the politics domain using public data collected from Twitter. The work has produced promising results. However, there are several limitations that need to be addressed and possible enhancements to be elucidated and marked as future work. It includes but is not limited to:

- Comprehensive ontologies being continuously updated by applying machine learning technologies i.e. driving data to obtain the domain knowledge (in reverse from the proposed approach),
- Analysing other social media such as Facebook, LinkedIn, and Weblogs, to name a few, and
- Incorporating the implementation and evaluation of the integration of domain-based trustworthiness in social business intelligence.

**Acknowledgements**

The paper has benefited from comments and suggestions by Professor Richard Baskerville, Department of Computer Information Systems, Georgia State University, and Dr Nik Thompson, Curtin Business School, Curtin University, the two reviewers, and the editor-in-chief, Professor Clyde Holsapple.

**Appendix: Glossary and Key Terms**

In order to make this article more understandable to the wide readership, we firstly provide a glossary and some key terms that are relevant in this paper.

*Ontology* is an explicit specification of a conceptualisation (Gruber, 1993). The specification is the representation of the conceptualisations in a concrete form (Stevens, 2001). The specification will lead to commitment in semantic structure. For example, categorisation of products and their features can be conceptualised into product ontology. The product ontology is then used for instance to share common understanding of the product taxonomies among people or software agents, to enable reuse of the product knowledge, and to make the product assumptions explicit.

*Entities* are concepts or things in the real world that is being modelled within the domain (Boyce & Pahl, 2007). Ontologies explicitly represent domains in the form of entities, properties, and relationships that exist in the real world and constitute the domain in focus. Entities are most likely to be nouns in sentences that describe the domain.

*Entity Type* is supertype or subtype in hierarchical order ("Entity Types and Supertypes,"). It is a term to denote that one is higher or lower in the hierarchy. The equivalent terms “parent" and "child" are also often used to define hierarchical order ("Ontology View,").

*Entity Extraction / Entity Recognition* is entity categorisation. The structure of the text can be analysed and parts / words of the sentences can be classified into categories for example person, location or organization (Alasiry, 2015).

*Controlled Vocabularies* are a complete list of terms being used by users and the domain experts. It typically includes preferred and variant terms and has a defined scope or describes a specific domain (Harpring, 2010).

*Concept Mapping* is a considered correlation of two different entities with the relation between entities defined via a specific property. It provides a means to capture concepts by constructing and refining the understanding of a domain (Leake, Maguitman, & Cañas, 2002).

*Taxonomy* is a hierarchy of concepts (only relation parent-child or subclass-superclass). Ontology has arbitrary complex relations between concepts other than concept hierarchy (Bai & Zhou, 2011).

*Text Classification* is the sorting of a set of documents into categories from a predefined set. Text classification attempts to determine whether the document discusses a given topic or contains a certain type of information (Cohen & Hersh, 2005).

*Terminology Extraction* is the extraction and identification of terms which are frequently used to refer to the concepts in a specific domain (Peñas, Verdejo, Gonzalo, & others, 2001).

*Relationship Extraction* is the extraction of many different semantics relationships between a pair of entities (Leng & Jiang, 2016).